\documentclass[aps,twocolumn,noshowpacs,superscriptaddress,floatfix]{revtex4-1}

\usepackage{quantikz}
\usepackage{amsmath,amssymb}
\usepackage[justification=justified]{caption}
\usepackage{xcolor}
\usepackage{verbatim}
\usepackage[normalem]{ulem}
\usepackage[rightcaption]{sidecap}
\usepackage{graphicx}
\usepackage{subfigure}
\usepackage{caption}
\usepackage[section]{algorithm} 
\usepackage{amsthm,algorithmic,epsfig,tikz}
\usepackage{float}
\usepackage{multirow}
\usepackage{hyperref}  
\hypersetup{colorlinks,citecolor=black,urlcolor=black,linkcolor =black,bookmarks=false,hypertexnames=true}

\def\bra#1{\mathinner{\langle{#1}|}}
\def\ket#1{\mathinner{|{#1}\rangle}}

\renewcommand{\part}[2]{\frac{\partial #1}{\partial #2}}

\begin{document}

\title{Complement Grover's Search Algorithm: An Amplitude Suppression Implementation}

\author{Andrew Vlasic}
\email{avlasic@deloitte.com}
\affiliation{Deloitte Consulting, LLP}

\author{Salvatore Certo}
\affiliation{Deloitte Consulting, LLP}

\author{Anh Pham}
\affiliation{Deloitte Consulting, LLP}

\date{\today}

\begin{abstract}
Grover's search algorithm was a groundbreaking advancement in quantum algorithms, providing a quadratic speed-up of querying for items. Since its creation, this algorithm it has been utilized in various ways, including in preparing specific states for the general circuit. However, as the number of desired items increases, so does the gate complexity of the sub-process that conducts the query. To counter this complexity, an extension of Grover's search algorithm is derived where the focus of the query is on the undesirable items in order to suppress the amplitude of the queried items. To understand its efficacy, this algorithm is implemented as a sub-process into the Quantum Approximate Optimization Algorithm (QAOA) and applied to a traveling salesman problem. For a basis of comparison, the results are compared against QAOA.  
\end{abstract}

\maketitle

\section{Introduction} \label{sec:intro}
Grover's search algorithm \cite{grover1996p} is a pivotal quantum routine that leads to quadratic speedups in an unstructured search, counting, and amplitude estimation procedures. Since its inception, this algorithm has been utilized as a form of state preparation, an essential subroutine in quantum algorithms. Specific examples of its use include the quantum topological analysis algorithm \cite{lloyd2016quantum} to prepare the boundary of the homology group, as well as Monte Carlo analysis \cite{rebentrost2018quantum}  for preparing states with various probability distributions. If only a subset of states are required for the circuit (see \cite{lloyd2016quantum}), there is potential for the depth complexity of the Grover subroutine to substantially increase, potentially eroding or eliminating the quadratic advantage it provides. 

This increase in complexity is found within the \textit{oracle query} that searches for the desirable states, and one may observe as the number of desirable states increases the gate complexity of the oracle increases. In more detail, the oracle is the first step in the algorithm and is constructed to add a negative phase to the desired states. Depending on the task at hand and the complexity of the oracle query, it may be more efficient, with regards to the  circuit requirements and gate depth, to invert this step, and instead to construct the oracle to add a positive phase to the undesirable solutions and a negative phase to the other states. Ergo, this oracle would search for fewer states, thereby decreasing complexity. This paper describes in detail how one may construct such an oracle. 

\section{An Overview of Grover's Search Algorithm} \label{sec:overview}
Grover's search algorithm works by locating items in unstructured data by increasing the amplitude of the desired items relative to the undesirable items. Therefore, once the circuit is measured, there is a higher probability of identifying the desirable items.

The algorithm is comprised of a periodic function that can be applied multiple times to iteratively change the amplitudes. In particular, the Grover iteration is a two step process, where first we define the oracle operator $O$ defined as $O:\ket{x}\ket{q} \to \ket{x}\ket{q \oplus f(x)}$, where $\oplus$ is the modulo 2 operation and $f(x) = 1$ if $x$ is the identified state and $0$ otherwise. One way to implement the oracle is by utilizing a Hadamard operator on the ancilla qubit, wherein the evolution of qubits including this ancilla has the form
$$
O\bigg( \ket{x}\Big( \frac{\ket{0} - \ket{1} }{ \sqrt{2} } \Big) \bigg) = (-1)^{f(x)} \ket{x} \Big( \frac{\ket{0} - \ket{1} }{ \sqrt{2} } \Big).
$$
The Hadamard state is crucial in seamlessly adding the local $-1$ phase shift to the desired items. 

To understand the oracle more concretely,consider a two qubit register and a one qubit ancilla with an oracle that identifies the state $\ket{11}$. The oracle would be the Toffoli gate controlled on the state $\ket{11}$ which, targeting Hadamard state, would add a negative phase. 

Once the desirable states are identified, there is an increase to their amplitudes using the \textbf{diffuser}, defined by the operator $D = 2 \ket{\psi} \bra{\psi} - I$.  Often referred to as an \textit{inversion about the mean}, the easiest way to implement the diffuser is to apply a phase shift of $-1$ to $\ket{0}$. One may observe the diffuser is of the form of a Householder matrix, and that $ 2 \ket{\psi} \bra{\psi} - I_{2^{n}}  = H^{\otimes n} \big( 2 \ket{0}\bra{0}^{\otimes n} - I_{2^{n}} \big)H^{\otimes n}$, since $H$ is self adjoint. 

\begin{figure}[ht]
\noindent   \resizebox{\columnwidth}{!}{
    \begin{quantikz}[thin lines] 
        \lstick{Register 0 $\ket{0}$} \gategroup[wires=4,steps=3,style={rounded corners,fill=cyan!10, inner xsep=2pt}, background]{{\sc Initial}} &\qw & \gate{H}  & \ctrl{1} \gategroup[wires=4,steps=1,style={rounded corners,fill=cyan!25, inner xsep=2pt}, background]{{\sc Oracle}} & \gate{H} \gategroup[wires=4,steps=5,style={rounded corners,fill=cyan!40, inner xsep=2pt}, background]{{\sc Diffuser}}    & \gate{X} &  \ctrl{1} & \gate{X} & \gate{H} &\qw \rstick[wires=3]{}
       \\ \lstick{Register 1 $\ket{0}$} &\qw & \gate{H} &\octrl{1} &\gate{H} & \gate{X} &  \ctrl{1} & \gate{X} & \gate{H}&\qw & |[meter]|
       \\ \lstick{Register 2 $\ket{0}$} &\qw & \gate{H} &\octrl{1} &\gate{H} & \gate{X} &  \ctrl{1} & \gate{X} & \gate{H}&\qw
       \\ \lstick{Ancilla $\ket{0}$}& \gate{X} & \gate{H} &\gate{X} &\qw & \qw &  \gate{X} & \qw & \qw&\qw
    \end{quantikz}
}
\caption{The circuit above displays a Grover's search algorithm for the string $\ket{001}$. One may use the open source quantum simulation package Qiskit \cite{aleksandrowicz2019qiskit} for computational verification that this circuit finds the state $\ket{001}$.}
\label{fig:clssl}
\end{figure}
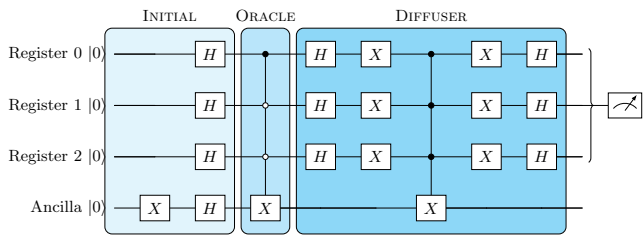

The subroutine with $\ket{0}^{\otimes n}\ket{0}$ qubits searching for the string $x_0$ is implemented as follows.
\begin{enumerate}
    \item Prepare the register qubits into the distribution state and prepare the ancilla qubit with the gate $H^{\otimes n} \otimes HX$, where $X$ is the Pauli operator. This gives the mathematical representation $\displaystyle \frac{1}{ \sqrt{2^n} } \sum_{x=0}^{2^n - 1} \ket{x} \Big( \frac{\ket{0} - \ket{1} }{ \sqrt{2} } \Big)$.  
    \item Apply the oracle $O$, which yields $\displaystyle \frac{1}{ \sqrt{2^n} } \sum_{x=0}^{2^n - 1} (-1)^{ \delta_{x_0} } \ket{x} \Big( \frac{\ket{0} - \ket{1} }{ \sqrt{2} } \Big)$. 
    \item Apply the adjoints of the Hadamard gates $H^{\otimes n}$, recalling $H$ is a self-adjoint operator.
    \item Apply conditional phase shift on all pure states of $-1$, except the state $\ket{0}^{\otimes (n)}$.
    \item Apply the Hadamard gates $H^{\otimes n}$.
    \item Repeat steps 2 through 5, as necessary, to approximate the state $\displaystyle \ket{x_0}\Big( \frac{\ket{0} - \ket{1} }{ \sqrt{2} } \Big)$.
    \item Finally, measure all the qubits, except for the ancilla qubit, in the computational basis. 
\end{enumerate}
Figure \ref{fig:clssl} gives an explicit example of this algorithm in the form of a circuit.

Since the Grover iteration may be repeated for several iterations in the circuit,the operators in steps 2-5 are known as the \textbf{Grover Operator}. One may see that the Grover operator, say $G$, has the form
$G = H^{\otimes n} \big( 2 \ket{0}\bra{0}^{\otimes n} - I_{2^{n}} \big)H^{\otimes n} O = \Big( 2 \ket{\psi} \bra{\psi} - I_{2^{n}} \Big) O$
where $\ket{\psi}$ is the uniformly distributed states of superpositions.

Although the tensor-ed Hadamard gate $H^{\otimes n}$ assumes a uniform distribution, this Grover operator subroutine has a natural extension for any sub-process that requires certain states to be fed downstream in the larger algorithm. One may see that any distribution may be consideredto evaluate  the sub-process, with an arbitrary distribution $\mathcal{A}$. having an overall Grover operator notation with  the general form  $\mathcal{A} \big(2\ket{0}\bra{0} - I \big)\mathcal{A}^{\dagger} O$. Other extensions of the algorithm include a creative extension of the Grover operator in the quantum analog of derivative pricing \cite{rebentrost2018quantum}. The authors in \cite{suzuki2020amplitude} also give a general extension of Grover's algorithm. An in-depth description of the Grover search algorithm can be found in Chapter 6 in the book \cite{nielsen2002quantum}. 

Lastly, it should be noted that because of the intricate nature of Grover's algorithm, its implementation on Noisy Intermediate Scale Quantum (NISQ) quantum processing units (QPUs) has made the algorithm with qubit size four or more unusable. However, the authors in \cite{zhang2022quantum} propose a method to bypass this shortcoming and enable a version of Grover's algorithm for NISQ devices.

\section{Grover's Amplitude Suppression Algorithm}\label{sec:comp} 
While the diffuser is fairly optimal and scales well with the dimension of the data, there is a potential to decrease the gate complexity of the oracle. There have been many advances from many different perspectives on Grover's algorithm \cite{yoder2014fixed, byrnes2018generalized,li2018complementary,cafaro2019continuous,gilliam2020optimizing}. In particular, the advancement in \cite{li2018complementary} focuses on utilizing sets within the search space, with the goal of decreasing the number of gates required. 

A decrease in gate complexity may come from changing the focus of the oracle from the set of  desired states to the set of undesirable state. To understand this perspective, we first denote the set $S$ as the collection of desirable pure states, $S^{\complement}$ as the set of undesirable states, $S \cup S^{\complement}$ as encompossing all pure states, and $|\cdot|$ as the cardinality of the set. Then,
\begin{equation*}
\begin{split}
\displaystyle  O & \bigg( \frac{1}{\sqrt{2^n}} \sum_{i=0}^{2^n-1}\ket{i} \left( \frac{\ket{0} -\ket{1} }{\sqrt{2}} \right) \bigg)  = \frac{1}{ \sqrt{2^n} } \sum_{i\in S} -\ket{i}\left( \frac{\ket{0} -\ket{1} }{\sqrt{2}} \right)\\ 
& + \frac{1}{ \sqrt{2^n} } \sum_{i' \in S^{\complement}}\ket{i'}\left( \frac{\ket{0} -\ket{1} }{\sqrt{2}} \right).
\end{split}
\end{equation*}

A potential method to decrease the complexity of the oracle is to focus on finding and labeling the undesirable states. While this may seem intuitive, it is not trivial to derive. 
Before explaining this algorithm in more detail, notations of the sets of desirable and undesirable states need to be defined.

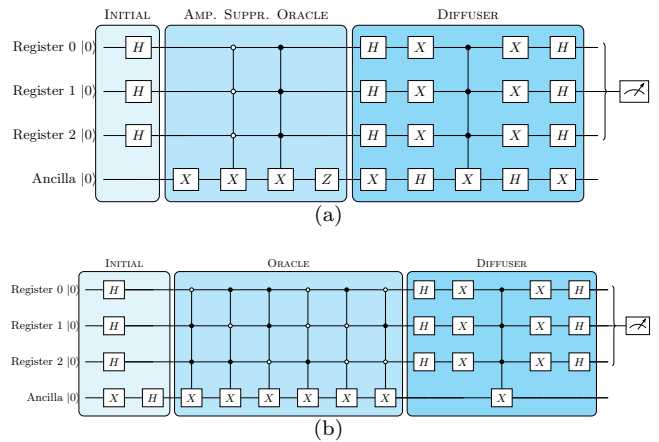
\begin{figure}[ht] 
 \subfigure[]{ \noindent   \resizebox{\columnwidth}{!}{  
    \begin{quantikz}[thin lines] 
        \lstick{Register 0 $\ket{0}$} \gategroup[wires=4,steps=2,style={rounded corners,fill=cyan!10, inner xsep=2pt}, background]{{\sc Initial}} & \gate{H}& \qw \gategroup[wires=4,steps=4,style={rounded corners,fill=cyan!25, inner xsep=2pt}, background]{{\sc Amp. Suppr. Oracle}}  & \octrl{1} & \ctrl{1}& \qw  & \gate{H} \gategroup[wires=4,steps=5,style={rounded corners,fill=cyan!40, inner xsep=2pt}, background]{{\sc Diffuser}}    & \gate{X} &  \ctrl{1} & \gate{X} & \gate{H} &\qw \rstick[wires=3]{}
       \\ \lstick{Register 1 $\ket{0}$} & \gate{H}& \qw &\octrl{1} & \ctrl{1} & \qw &\gate{H} & \gate{X} &  \ctrl{1} & \gate{X} & \gate{H}&\qw & |[meter]|
       \\ \lstick{Register 2 $\ket{0}$} &  \gate{H}& \qw &\octrl{1} & \ctrl{1} & \qw &\gate{H} & \gate{X} &  \ctrl{1} & \gate{X} & \gate{H}&\qw
       \\ \lstick{Ancilla $\ket{0}$}& \qw &\gate{X} &\gate{X} &\gate{X} &\gate{Z} &\gate{X} & \gate{H} &  \gate{X} & \gate{H} & \gate{X}&\qw
    \end{quantikz}
} }
\subfigure[]{ \noindent   \resizebox{\columnwidth}{!}{    
    \begin{quantikz}[thin lines] 
        \lstick{Register 0 $\ket{0}$} \gategroup[wires=4,steps=3,style={rounded corners,fill=cyan!10, inner xsep=2pt}, background]{{\sc Initial}} & \gate{H} & \qw & \octrl{1} \gategroup[wires=4,steps=6,style={rounded corners,fill=cyan!25, inner xsep=2pt}, background]{{\sc Oracle}} & \ctrl{1} & \ctrl{1} & \octrl{1} & \ctrl{1} & \octrl{1} &  \gate{H} \gategroup[wires=4,steps=5,style={rounded corners,fill=cyan!40, inner xsep=2pt}, background]{{\sc Diffuser}}    & \gate{X} &  \ctrl{1} & \gate{X} & \gate{H} &\qw \rstick[wires=3]{}
       \\ \lstick{Register 1 $\ket{0}$} & \gate{H} & \qw      &  \ctrl{1}   & \octrl{1} & \ctrl{1} & \octrl{1} & \octrl{1} & \ctrl{1} &\gate{H} & \gate{X} &  \ctrl{1} & \gate{X} & \gate{H}&\qw & |[meter]|
       \\ \lstick{Register 2 $\ket{0}$} & \gate{H} & \qw      &   \ctrl{1}  & \ctrl{1} & \octrl{1} & \ctrl{1} & \octrl{1} & \octrl{1} &\gate{H} & \gate{X} &  \ctrl{1} & \gate{X} & \gate{H}&\qw
       \\ \lstick{Ancilla $\ket{0}$}    & \gate{X} & \gate{H} &\gate{X}     & \gate{X} & \gate{X} &\gate{X} & \gate{X} & \gate{X} & \qw & \qw &  \gate{X} & \qw & \qw & \qw
    \end{quantikz}
} }
\caption{(a) This circuit displays an amplitude suppression of Grover's search algorithm for all states except for $\ket{000}$ and $\ket{111}$. For the six desirable states, there are two oracles and two single qubit gates, instead of the implied six oracles. (b) This circuit displays Grover's search algorithm for the six desirable states}
\label{fig:comp}
\end{figure}

To obtain the goal of adding the negative phase to the desirable states, an ancilla qubit with state $\ket{0}$ is introduced. In the oracle, the ancilla is first flipped to the state $\ket{1}$ with the Pauli $X$ gate, which allows the Pauli $Z$ gate to give the negative phase to all desired states. If a state is an undesirable state, the ancilla qubit is flipped back to $\ket{0}$. Unlike the standard oracle, the ancilla qubit is not put into a Bell state. Although having an ancilla qubit without superposition may be perceived as losing an advantage, there are instances of applying control gates to pure state ancilla \cite{vazquez2021efficient, suzuki2020amplitude}.  

After setting the qubits in the register into a uniform superposition and applying amplitude suppression, the oracle provides superpositions, which can be shown mathematically as
\begin{equation*}
\begin{split}
\displaystyle O_{S} & \bigg( \frac{1}{\sqrt{2^n}}\sum_{i=1}^{2^n-1}\ket{i}\ket{0} \bigg) = \frac{1}{\sqrt{2^n}}\sum_{i\in S }-\ket{i}\ket{1} \\
& + \frac{1}{\sqrt{2^n}}\sum_{i' \in S^{\complement}}\ket{i'}\ket{0}.  
\end{split}
\end{equation*}

In order to use this complement, oracle, the standard diffuser operator (see Figure \ref{fig:comp}.b) requires adjustment, since the operator is expecting a Bell state in the ancilla register. Therefore, the addition of the Pauli-$X$ gate followed by the Hadamard gate is included in the diffuser. Since the complement oracle requires a pure state, the inverse of these gates are included after the general qubit Toffoli gate; see Figure \ref{fig:comp}.a for an example circuit.  

The general process of the algorithm is as follows. 
\begin{enumerate}
    \item Place the register qubits into the distribution state using the gate $H^{\otimes n}\otimes I$ , and add an ancilla qubit initialized to the $\ket{0}$ state, leading to $\displaystyle \frac{1}{\sqrt{2^n}}\sum_{i=0}^{2^n-1}\ket{i}\ket{0}$.
    \item Apply the oracle $O_{S}$.
    \item Apply Hadamard gates $H^{\otimes n}$.
    \item Apply conditional phase shift on all states, except the state $\ket{0}^{\otimes (n+1)}$, to perform the phase shift of $-1$.
    \item Apply the Hadamard $H^{\otimes n}$.
    \item Repeat steps 2 through 5, when necessary, to approximate the final state $\displaystyle \frac{1}{\sqrt{|S|}}\sum_{i\in S}\ket{i}\ket{1}$.
    \item Finally, measure all the qubits, except for the ancilla qubit, in the computational basis.
\end{enumerate}

\begin{figure}[ht]
    \centering
    \subfigure[]{\includegraphics[width=0.45\textwidth]{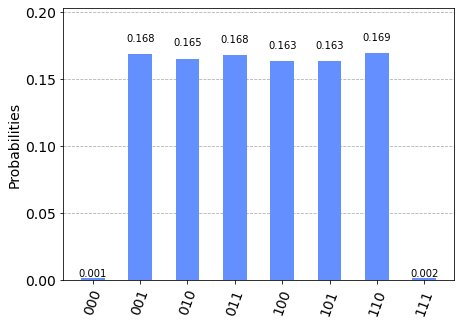}} 
    \subfigure[]{\includegraphics[width=0.45\textwidth]{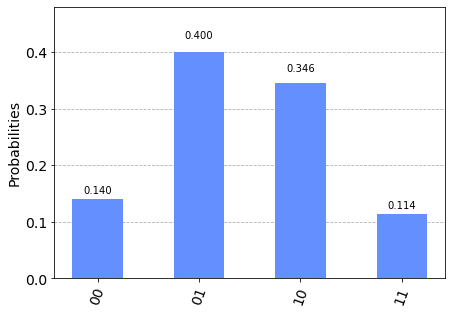}} 
    \caption{The bar graphs are results from running the algorithm on a simulator and a QPU. Figure (a) displays the results from the circuit in Figure \ref{fig:comp} ran on Qiskit \cite{aleksandrowicz2019qiskit} using a simulator backend after 3 Grover Iterations. Figure (b) are the results of running the circuit in Figure \ref{fig:comp} but with a two qubit register and ran on an IBM QPU.}
    \label{fig:runs_qpu}
\end{figure}

To make this algorithm more tangible, Figure \ref{fig:comp}(a) shows a circuit of preparing the states $\ket{001},\ldots,\ket{110}$. Figure \ref{fig:runs_qpu} displays results of this algorithm with registers of size three qubits and two qubits. 

One may observe that the action of the adjusted-diffuser acts in the same manner as the standard diffuser. Moreover, while there is a subtle difference in the ancilla qubit, the register for the operator $O_{S}$ is the same as $O$. Therefore, the original analysis on the number of iterations needed still holds true; see \cite{nielsen2002quantum} for further information. In particular, for an $R$ number of iterations, an $N$ total number of states, and an $M$  number of undesirable states, the inequality  $\displaystyle R \leq \Big\lceil \frac{N}{M} \Big\rceil$ holds, where $\lceil \cdot \rceil$ is the ceiling function.

\begin{figure*}[t!]
    \centering
    \subfigure[]{\includegraphics[width=0.4\textwidth]{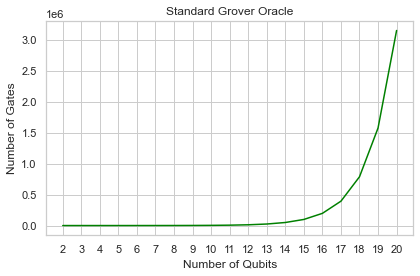}} 
    \subfigure[]{\includegraphics[width=0.4\textwidth]{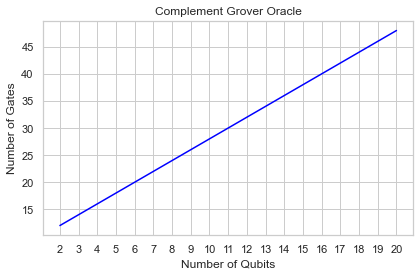}} 
    \subfigure[]{\includegraphics[width=.4\textwidth]{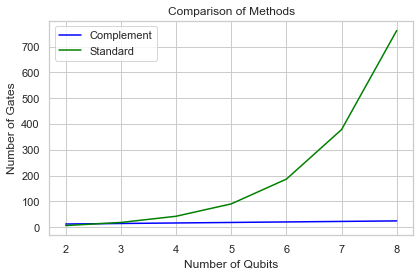}}
    \caption{The figures show the increase in the number of gates required for the oracle in the circuit in Figure \ref{fig:comp} compared to the classical analog this this circuit. Figure (a) is the standard oracle. Figure (b) is oracle size in Figure \ref{fig:comp}. Finally, Figure (c) compares these different methods.}
    \label{fig:growths}
\end{figure*}

\section{Analysis of the Oracle Depth}\label{sec:anal} 
\begin{figure}[ht] 
\noindent   \resizebox{200pt}{50pt}{
    \begin{quantikz}[thin lines] 
       \lstick{Register 0 $\ket{0}$} & \qw & \ctrl{1} & \qw & \qw 
       \\ \lstick{Register 1 $\ket{0}$} & \gate{X}& \ctrl{1} & \gate{X} & \qw 
       \\ \lstick{Register 2 $\ket{0}$} & \qw & \ctrl{1} & \qw & \qw 
       \\ \lstick{Ancilla $\ket{0}$}& \qw &\gate{X} &\qw &\qw
    \end{quantikz}
}
\caption{The circuit displays an oracle identifying the state $\ket{101}$.}
\label{fig:orc_exmpl}
\end{figure}
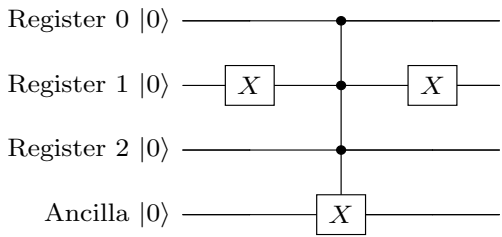

While there are creative ways to prepare states \cite{huang2018demonstration}, the complement oracle based algorithm presented above is scalable, even given the increasing gate count. Hence,  complemenet oracle based Grover's search is agnostic to the number of qubits required. Furthermore, the addition of the Pauli $X$ gate and Pauli $Z$ gate stays constant as the number of qubits in the register grows. Therefore, with at least a two qubit register, the subtraction of one control gate in the oracle yields a more efficient circuit.

For a deeper understanding of the potential increased efficiency of this new algorithm, the circuit depth of the standard oracle and complement oracle will be compared. The oracles from the classical Grover's search algorithm are assumed to be scalable, bypassing creative oracles that are qubit-number specific. To be scalable for each string in the desirable state, there is a general Toffoli gate identifying a particular state with appropriate Pauli $X$ gates to flip the qubit then flip back to the original state; see Figure \ref{fig:orc_exmpl} for an example. This particular implementation of flagging states should be hardware agnostic.

Each Pauli $X$ gate and each general Toffoli gate are counted as one. The reason a general Toffoli gates is counted as one is to adjust for the different decomposition of native gates that this operator would be implemented with on the respective QPUs. The diffusion operators are taken to be the same in order to compare the same algorithms, giving a fixed additional four gates within standard Grover's algorithm.

The analysis is of the circuit depth is in Figure \ref{fig:growths}. Taking the register size of 2 to 20 qubits, the number of operators in each oracle is counted. Since Grover's algorithm prepares all states except for $\ket{0\ldots0}$ and $\ket{1\ldots 1}$ the growth of the required state preparations in the classical implementation is $2^n-2$ for $n$ qubits, while the amplitude suppression implementation always requires $2$ states to prepare. In particular, the standard case grows with $n$ the following formula $\displaystyle  (2^n-2) + 2\cdot \sum_{i=1}^{n-1} \left( \begin{array}{c} n  \\ i \end{array} \right)$, and the amplitude suppression case grows as $8 + 2\cdot n$. From this one may observe the standard implementation grows exponentially and amplitude suppression grows linearly. Figure \ref{fig:comp} shows and compares these different growths.

While this is an extreme case to explore, there are examples which such an algorithm is not required. For example, if the first of position needs to be in the state $\ket{1}$ for a number of particular states. In this instance, one may implement Grover around this position with a simple oracle for state preparation. However, this analysis displays the potential increase in efficiency when intricate states are required in the general algorithm.

\section{Application to the Quantum Approximate Optimization Algorithm}\label{sec:qaoa}
On application of this algorithm is to use it as state preparate for the Quantum Approximate Optimization Algorithm (\textbf{QAOA}) \cite{farhi2014quantum}, which was derived as a means to find the lowest energy within in a system. This was achieved by taking the adiabatic quantum evolution and applying the Trotter product formula to approximate the Hamiltonian. The ground state is then mapped to the solution of the mathematical problem that was encoded into the circuit. From the derivation, one may see how numerous combinatorial optimization problems can be translated into a QAOA circuit, including various graph problems and various mathematical programming problems. 

QAOA requires an ansatz for the initial condition and typically requires application of a classical optimizer to derive the parameters that yield the optimal or near-optimal solution. A uniform distribution of all the states was usually taken as the initial sub-process. However, there have been recent advancements that utilize the mixer Hamiltonian to guide the derivation of superpositions closer to the optimal solution.  \cite{egger2021warm}. This algorithm is noted as the warm start quantum approximate optimization algorithm (\textbf{WS-QAOA}). For completeness, other derivations using custom mixer Hamiltonian for QAOA have been considered \cite{tate2021classically}.  

WS-QAOA works by relaxing the QUBO and solving it classically to get an approximate solution.  This approximate solution is then given as an initial state to QAOA.  For problems where the gap of the relaxation to the QUBO solution is minimal, this method is effective and efficient.  When this gap is adequately large however, its efficacy is drastically reduced.  Some instances where a relaxed problem solution may be far from the true QUBO minimum is in cases like the Traveling Salesman Problem.  

\begin{figure}[h]
    \centering
    \includegraphics[width=229px]{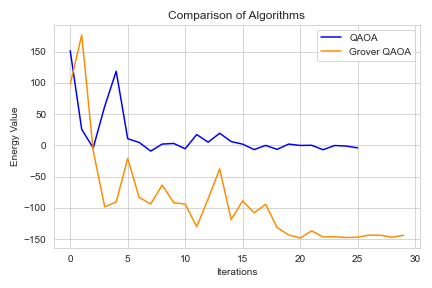}
    \caption{Comparison of the QAOA and QAOA using amplitude suppression Grover as the initial start of the circuit.}
    \label{fig:qaoa}
\end{figure}

To show the efficacy the algorithm in Figure \ref{fig:comp} (a) is implemented as the first sub-process in QAOA and applied to a toy traveling salesman problem. In order to provide effective comparison, the same traveling salesman problem is applied to QAOAand the results are compared. Comparison against this initialization method against others methods, like WS-QAOA, is left for future research. The circuits are written in Qiskit \cite{aleksandrowicz2019qiskit}, and as noted in \cite{egger2021warm}, there are packages to run the original QAOA, (as well as WS-QAOA). The results are given in Figure \ref{fig:qaoa} where the original QAOA using amplitude suppression Grover's algorithm out-performs Qiskit's QAOA package. The reason for the performance boost is that the amplitude suppression Grover's algorithm decreases the amplitudes of the infeasible states, allowing for initial higher probabilities of feasible states, in which the solution can be found faster.

\section{Discussion and Conclusion}
In the case where the desirable states $S^{C}$ outnumber the undesirable states $S$, the algorithm presented above represents an efficient way to structure the oracle. While this structure normally depends entirely on the domain and query involved in the algorithm, certain applications of state preparation include specific states to be omitted, which prompts a different approach to efficiently construct the Grover operator.  The benefit proposed here relies on the fact that we are only adding an ancilla qubit within the context of decreasing the amplitudes of certain states, rather than during the initial preparation of the state.  In fact, some tasks that require the Grover Operator will see efficiency gains through this method structuring the oracle on the undesirable states, rather than the desired solutions themselves.

\section{Disclaimer}
About Deloitte: Deloitte refers to one or more of Deloitte Touche Tohmatsu Limited (“DTTL”), its global network of member firms, and their related entities (collectively, the “Deloitte organization”). DTTL (also referred to as “Deloitte Global”) and each of its member firms and related entities are legally separate and independent entities, which cannot obligate or bind each other in respect of third parties. DTTL and each DTTL member firm and related entity is liable only for its own acts and omissions, and not those of each other. DTTL does not provide services to clients. Please see www.deloitte.com/about to learn more.

Deloitte is a leading global provider of audit and assurance, consulting, financial advisory, risk advisory, tax and related services. Our global network of member firms and related entities in more than 150 countries and territories (collectively, the “Deloitte organization”) serves four out of five Fortune Global 500® companies. Learn how Deloitte’s
approximately 330,000 people make an impact that matters at www.deloitte.com. 
This communication contains general information only, and none of Deloitte Touche Tohmatsu Limited (“DTTL”), its global network of member firms or their related entities (collectively, the “Deloitte organization”) is, by means of this communication, rendering professional advice or services. Before making any decision or taking any action that
may affect your finances or your business, you should consult a qualified professional adviser. No representations, warranties or undertakings (express or implied) are given as to the accuracy or completeness of the information in this communication, and none of DTTL, its member firms, related entities, employees or agents shall be liable or
responsible for any loss or damage whatsoever arising directly or indirectly in connection with any person relying on this communication. 
Copyright © 2022. For information contact Deloitte Global.

\bibliographystyle{unsrt}
\bibliography{GroverBib}
\end{document}